\documentclass[twocolumn,preprintnumbers,amsmath,amssymb,aps,prb]{revtex4}
\usepackage{graphicx}
\begin{document}

\title{
Collective Transport for Active Matter Run and Tumble Disk Systems on a Traveling Wave Substrate
} 
\author{
Cs. S\'{a}ndor$^{1,2}$, A. Lib\'{a}l$^{1,2}$, C. Reichhardt$^{1}$ and C. J. Olson Reichhardt$^{1}$}
\affiliation{$^{1}$Theoretical Division and Center for Nonlinear Studies,
Los Alamos National Laboratory, Los Alamos, New Mexico 87545, USA}
\affiliation{$^{2}$Mathematics and Computer Science Department, Babe{\c s}-Bolyai University, Cluj, Romania 400081}

\date{\today}
\begin{abstract}

  We numerically examine the transport of an assembly of active 
  run-and-tumble disks
interacting with a traveling wave substrate.
We show that as a function of substrate strength, wave speed, disk activity, and
disk density, a variety of dynamical phases arise
that are correlated with the structure and net flux of disks.
We find that there is a sharp transition
into a state where the disks are only partially coupled to the substrate and form
a phase separated cluster state.
This transition is associated with a drop in the net disk flux
and
can occur as a function
of the substrate speed, maximum substrate force,
disk run time, and disk density.
Since variation of the disk activity parameters produces different disk
drift rates for a fixed traveling wave speed on the substrate,
the system we consider
could be used as an efficient method for active matter species separation.
Within the cluster phase,
we find that in some regimes the motion of the cluster center of mass is in the opposite
direction to that of the traveling wave, while when the maximum substrate force
is increased,
the cluster drifts in the direction of the traveling wave.
This suggests that swarming or clustering motion can serve as a method by which
an active system can collectively move against an external drift.
\end{abstract}
\maketitle

\vskip2pc

\section{Introduction}

Collections of interacting self-motile objects
fall into the class of systems known as active
matter \cite{1,2},
which can be biological in nature
such as swimming bacteria \cite{3} or animal herds \cite{4},
a social system such as pedestrian or traffic flow \cite{5},
or a robotic swarm \cite{6,7}.
There are also a wide range of artificial active matter
systems such as self-propelled colloidal particles \cite{8,9,10}.
Studies of these systems have  generally focused on the case
where the motile objects interact with either a  smooth or
a static substrate; however, the field is now advancing to a point where
it is possible to ask
how such systems behave
in more complex static or dynamic environments.

One subclass of active systems
is a collection of interacting disks that undergo either run-and-tumble \cite{11,12} or
driven diffusive \cite{13,14,15} motion.
Such systems
have been shown to exhibit a transition
from a uniform density liquid state
to a motility-induced phase separated state
in which the disks form dense clusters surrounded
by a low density gas phase \cite{9,10,11,12,13,14,15,16,17,18}.
Recently it was shown that when phase-separated run-and-tumble disks are
coupled to a random pinning substrate, a transition to a uniform density liquid state
occurs as a function of the maximum force exerted by the substrate \cite{19}.
In other studies of run-and-tumble disks driven over an obstacle array by a dc driving force,
the onset of clustering coincides with a drop in the net disk transport since a large cluster
acts like a rigid object that can only move through the obstacle array with difficulty; in
addition, it was shown that the disk transport was maximized at an optimal
activity level or disk running time \cite{20}.
Studies of flocking or swarming disks that obey modified Vicsek models of
self-propulsion  \cite{21}
interacting with obstacle arrays
indicate 
that there is an optimal
intrinsic noise level at which collective swarming occurs \cite{22,23},
and that transitions between swarming and non-swarming states can occur as
a function of increasing substrate disorder \cite{24}.
The dynamics in such swarming models differ from those of
the active disk systems, so it is not clear whether the  same behaviors will occur across
the two different systems.

A number of studies have already considered
active matter such as bacteria or run-and-tumble disks
interacting with periodic obstacle arrays \cite{25} or
asymmetric arrays  \cite{26,27,28,29}.
Self-ratcheting behavior occurs for the asymmetric arrays
when the combination of broken detailed balance and the substrate asymmetry
produces
directed or ratcheting motion of the active matter particles
\cite{30,31},
and it is even possible to couple passive particles to the active matter particles
in such arrays in order to shuttle cargo across the sample \cite{29}.
In the studies described above, the substrate is static, and external driving
is introduced via fluid flow or chemotactic effects; however, it is also possible
for the substrate itself to be dynamic, such as in the case of
time dependent optical traps  \cite{32,33} or a traveling wave substrate.
Theoretical and experimental studies of colloids in traveling wave potentials
reveal a rich variety of dynamical phases,
self-assembly behaviors, and directed transport \cite{34,35,36,37,38,39,40}.

Here we examine a two-dimensional system of run and tumble active
matter disks that can exhibit motility induced phase separation
interacting with a periodic quasi-one dimensional (q1D) traveling wave substrate.
In the low activity limit, the substrate-free system forms a
uniform liquid state, while in the presence of a substrate,
the disks are readily trapped by the substrate minima
and swept through the system by the traveling wave.
As the activity increases, a partial decoupling transition of the disks and the substrate
occurs, producing a drop in the net effective transport.  This transition is correlated
with the onset of the phase separated state,
in which the clusters act as large scale composite objects that cannot be transported
as easily as individual disks by the traveling wave.
We also find that the net disk transport is optimized at particular
traveling wave speeds, disk run length, and substrate strength.
In the phase separated state we observe an interesting effect where
the center of mass of each cluster moves in the direction opposite to that in which
the traveling wave is moving, and we also find reversals
to states in which the clusters and the traveling wave move in the same direction.
The reversed motion of the clusters arises due to asymmetric growth and shrinking rates
on different sides of the cluster.
The appearance of backward motion of the cluster center of mass
suggests that certain biological or social active systems can move against biasing
drifts by forming large collective objects or swarms.

\begin{figure}
\includegraphics[width=3.5in]{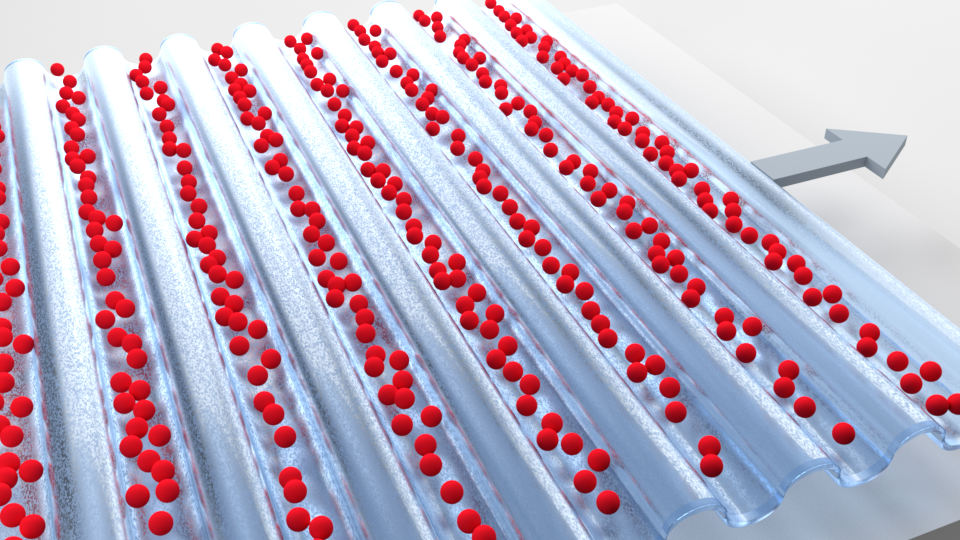}
\caption{Schematic of the system.
  Red spheres represent the active run and tumble disks in a two-dimensional system
  interacting with a periodic q1D traveling wave potential
  which is moving in the positive $x$-direction (arrow) with a wave speed of $v_{w}$.
}
\label{fig:1}
\end{figure}

\section{Simulation}
We model a two-dimensional system of $N$ run and tumble disks
interacting with a q1D traveling wave periodic substrate,
as shown in the schematic in Fig.~\ref{fig:1} where the substrate
moves to the right at a constant velocity $v_{w}$.
The dynamics of each disk is governed by the following overdamped equation of motion:
\begin{equation}
\eta \frac{d{\bf r}_i}{dt} = {\bf F}^{\rm inter}_i + {\bf F}^m_i + {\bf F}^{s}_i ,
\end{equation}
where the damping constant is $\eta = 1.0$.
The disk-disk repulsive interaction force ${\bf F}^{\rm inter}_i$
is modeled as a harmonic spring,
${\bf F}^{\rm inter}_i=\sum_{j\neq i}^N\Theta(d_{ij}-2R)k(d_{ij}-2R){\bf \hat d}_{ij}$,
where $R=1.0$ is the disk radius, $d_{ij}=|{\bf r}_i-{\bf r}_j|$ is the distance between
disk centers, ${\bf \hat d}_{ij}=({\bf r}_i-{\bf r}_j)/d_{ij}$, and the spring constant
$k=20.0$
is large enough to prevent significant disk-disk overlap under the conditions we
study
yet small enough to permit a computationally efficient time step
of $\delta t=0.001$ to be used.
We consider a sample of size $L \times L$ with $L=300$, and describe the disk density in
terms of the area coverage
$\phi = N\pi R^2/L^2$.
The run and tumble self-propulsion is modeled with a motor force  ${\bf F}^{m}_i$
of fixed magnitude $F^m=1.0$ that acts in a randomly chosen direction during
a run time of $\tilde{t}_{r}$.
After this run time, the motor force instantly reorients into a new
randomly chosen direction for the next run time.
We take $\tilde{t}_r$ to be uniformly distributed over the range
[$t_r,2t_r$], using run times ranging from $t_r=1 \times 10^3$ to $t_r=3 \times 10^5$.
For convenience we describe 
the activity in terms of the run length $r_l=F^mt_r\delta t$,
which is the distance a disk would move during a single run time
in the absence of a substrate or other disks.
The substrate is modeled as a time-dependent
sinusoidal force $F^s_i(t)=A_{s}\sin(2\pi x_i-v_w t)$
where $A_s$ is the substrate strength and $x_i$ is the $x$ position of disk $i$.  We take
a substrate periodicity of $a = 15$ so that the system contains
$20$ minima.
The substrate travels at a constant velocity of $v_{w}$ in the positive $x$-direction.
We measure the average drift velocity of the disks in the direction of the traveling wave,
$\langle V\rangle = N^{-1}\sum^{N}_{i}{\bf v}_{i}\cdot {\hat {\bf x}}$.
We  vary the run length, substrate strength, disk density, and wave speed.
In each case   
we wait for a fixed time
of $5 \times 10^6$ simulation time steps
before taking measurements to avoid any transient effects.

\begin{figure}
\includegraphics[width=3.5in]{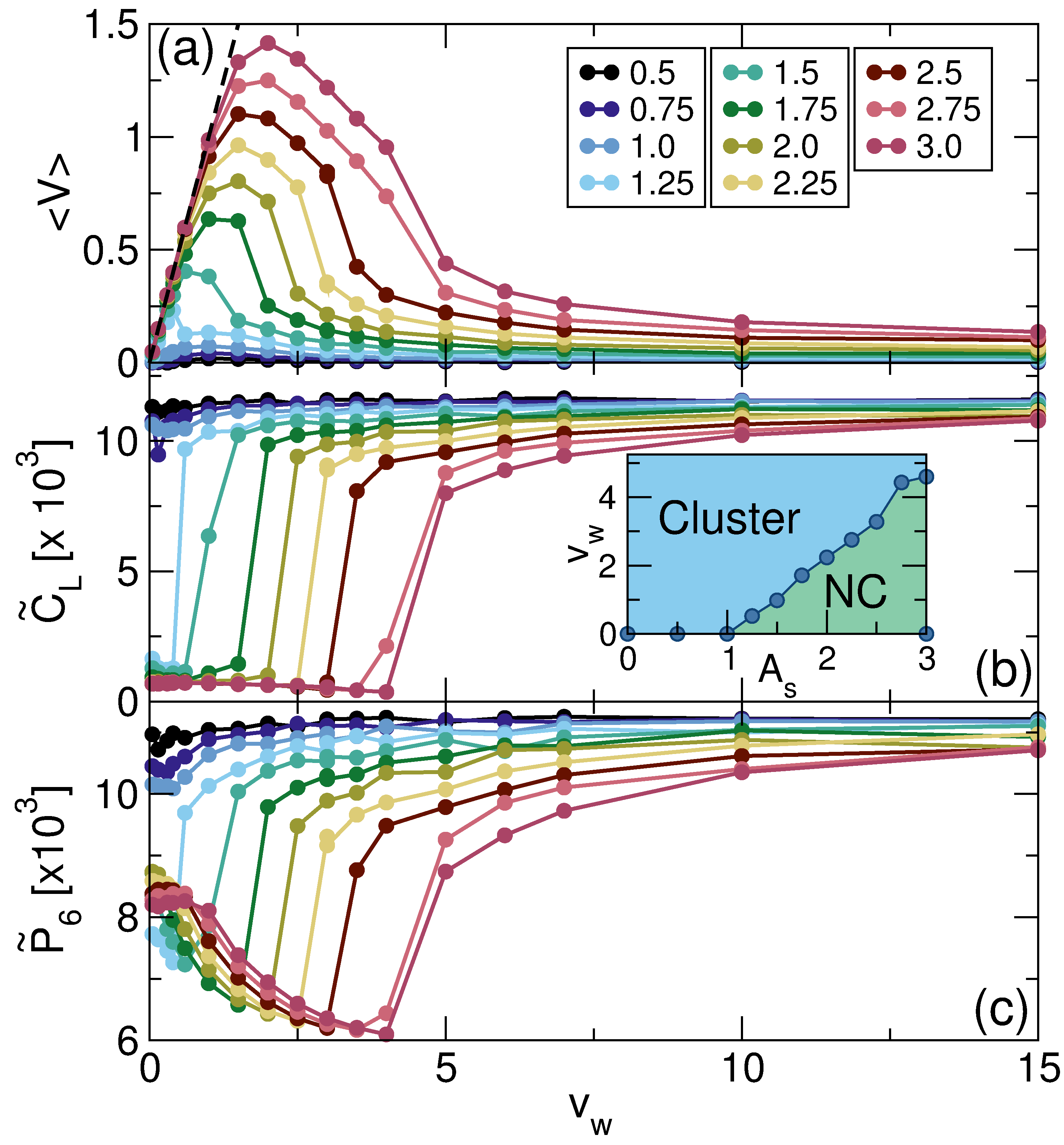}
\caption{(a) The average velocity per disk $\langle V\rangle$
  vs wave speed $v_{w}$ for a system with
  $N=13000$ disks, $\phi  = 0.45376$, and
  $r_l = 300$ for varied substrate strengths of $A_s=0.5$ to 3.0.
  The dashed line indicates the limit in which all the disks move at the wave
  speed, $\langle V\rangle = v_{w}$.
  (b) The corresponding number $\tilde{C}_L$ of disks that are in a cluster
vs wave speed. 
The inset shows the regions of cluster and non-cluster (NC) states
as a function of $v_{w}$ vs $A_{s}$. 
(c) The number $\tilde{P}_6$  of sixfold-coordinated
disks vs wave speed $v_w$.
}
\label{fig:2}
\end{figure}

\section{Results}
In Fig.~\ref{fig:2}(a) we plot the average velocity per disk
$\langle V\rangle$ versus wave speed $v_{w}$ at different substrate strengths
$A_{s}$ for a system containing $N=13000$ active disks,
corresponding to $\phi = 0.45376$,
at $r_l = 300$, a running length at which the substrate-free system
forms a phase separated state.
The number of disks that are in
the largest cluster, $\tilde{C}_L$, serves as an effective measure
of whether the system is in a phase separated state or not.  We measure $\tilde{C}_L$
using the cluster identification algorithm described in Ref.~\cite{Hermann}, and call
the system phase separated when $\tilde{C}_L/N>0.55$.  In Fig.~\ref{fig:2}(b) we plot
$\tilde{C}_L$ versus $v_w$ at varied $A_s$,
and in Fig.~\ref{fig:2}(c) we show the corresponding number of sixfold-coordinated
disks, $\tilde{P}_6=\sum_i^N\delta(z_i-6)$, where $z_i$ is the coordination number of
disk $i$ determined from a Voronoi construction \cite{cgal}.  In phase separated states,
most of the disks within a cluster have $z_i=6$ due to the triangular ordering of the
densely packed state.
In Fig.~\ref{fig:2}(a), the linearly increasing dashed line denotes
the limit in which all the disks
move with the substrate so that $\langle V\rangle = v_{w}$ .
At $A_{s} = 3.0$, $\langle V\rangle$ initially increases linearly, following the dashed
line, up to $v_{w} = 1.25$, 
indicating that there is a complete locking of the
disks to the substrate.
For $v_w > 1.25$, there is a slipping process in which the disks
cannot keep up with the traveling wave and jump to the next well.
A maximum in $\langle V\rangle$ appears near $v_{w} = 2.0$,
and there is a sharp drop in $\langle V\rangle$ near $v_{w} = 5.0$,
which also coincides with a sharp increase in  $\tilde{C}_L$ and $\tilde{P}_{6}$.
The  $\langle V\rangle$ versus $v_{w}$ curves for $A_{s} > 1.0$
all show similar trends, with a sharp drop
in $\langle V\rangle$ accompanied by an increase in
$\tilde{C}_{L}$ and $\tilde{P}_{6}$, showing that the onset of clustering
results in a sharp decrease in $\langle V\rangle$.
For $A_{s} \leq 1.0$, the substrate is weak enough that the system
remains in a cluster state even at $v_{w} = 0$,
indicating that a transition from a cluster to a non-cluster state
can also occur as a function of substrate strength.
In the inset of Fig.~\ref{fig:2}(b) we show the regions in which
clustering and non-clustering states appear as a function of $v_w$ versus $A_s$.
At $v_{w} = 0$, there is a substrate-induced transition from a
cluster to a non-cluster state near $A_{s} = 1.0$,
while for higher $A_s$, the location of the transition shifts linearly to higher $v_w$
with increasing $A_s$.
Since the motor force is $F^m=1.0$,
when $A_{s} < 1.0$ individual disks
can escape from the substrate minima,
so provided that $r_{l}$ is large enough, the disks can freely move
throughout the entire system and form a cluster state.
For $A_{s} > 1.0$, the disks are confined by the substrate minima,
but when $v_w$ becomes large enough, the disks can readily escape
the minima and again form a cluster state.

\begin{figure}
\includegraphics[width=3.5in]{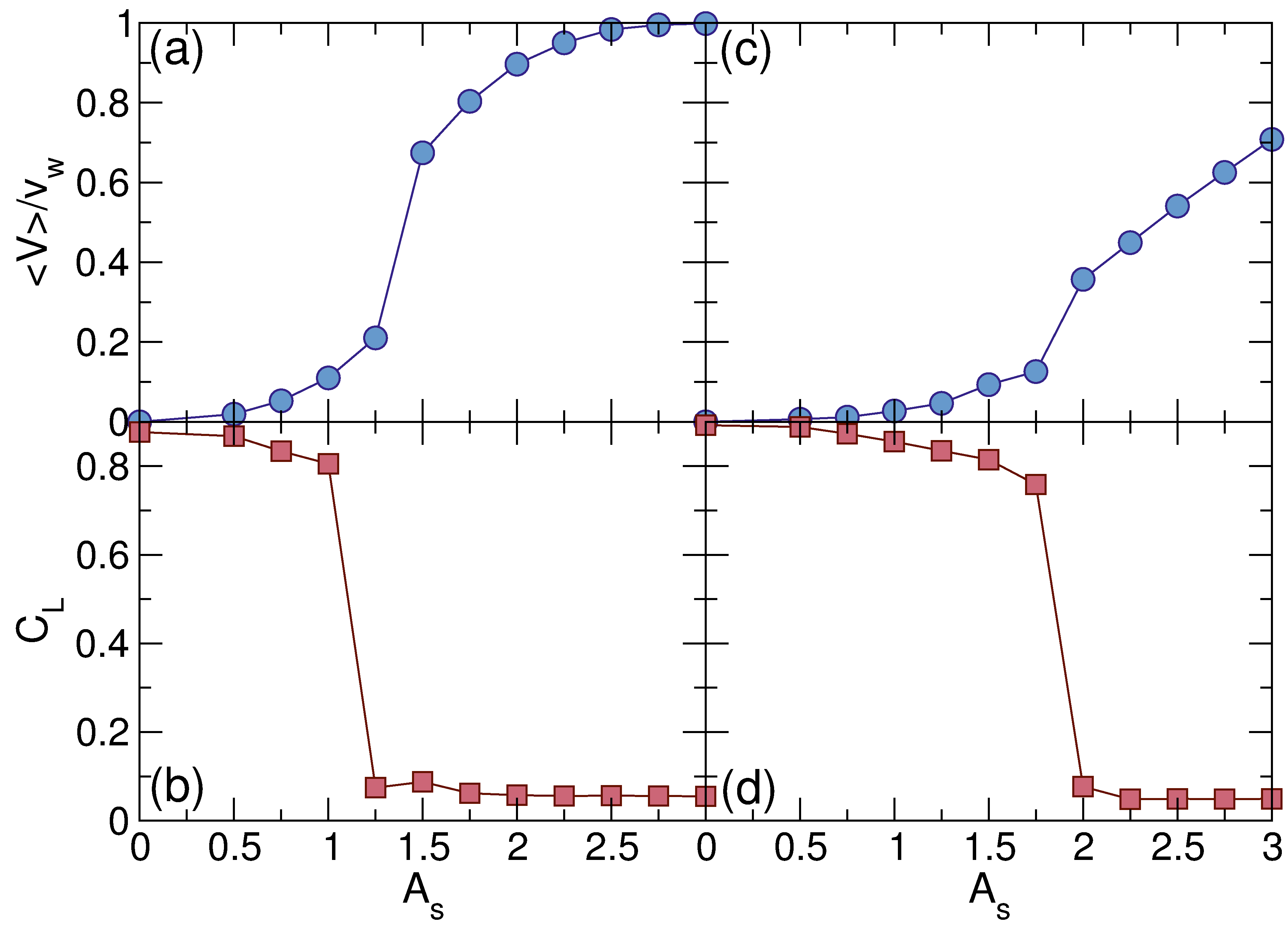}
\caption{(a) $\langle V\rangle/v_{w}$
  vs $A_{s}$ for the system in Fig.~\ref{fig:2}
  at $v_{w} = 0.6$.
  (b) The corresponding normalized $C_{L}$ showing
  that the  transition from a cluster to a non-cluster state
  coincides with an increase in $\langle V\rangle/v_w$.
  (c) $\langle V\rangle/v_{w}$ vs $A_s$ for the same system with $v_w=2.0$
  where the cluster to non-cluster transition occurs at a higher value of $A_{s}$.
  (d) The corresponding normalized $C_{L}$ vs $A_s$.
}
\label{fig:3}
\end{figure}

\begin{figure}
\includegraphics[width=3.5in]{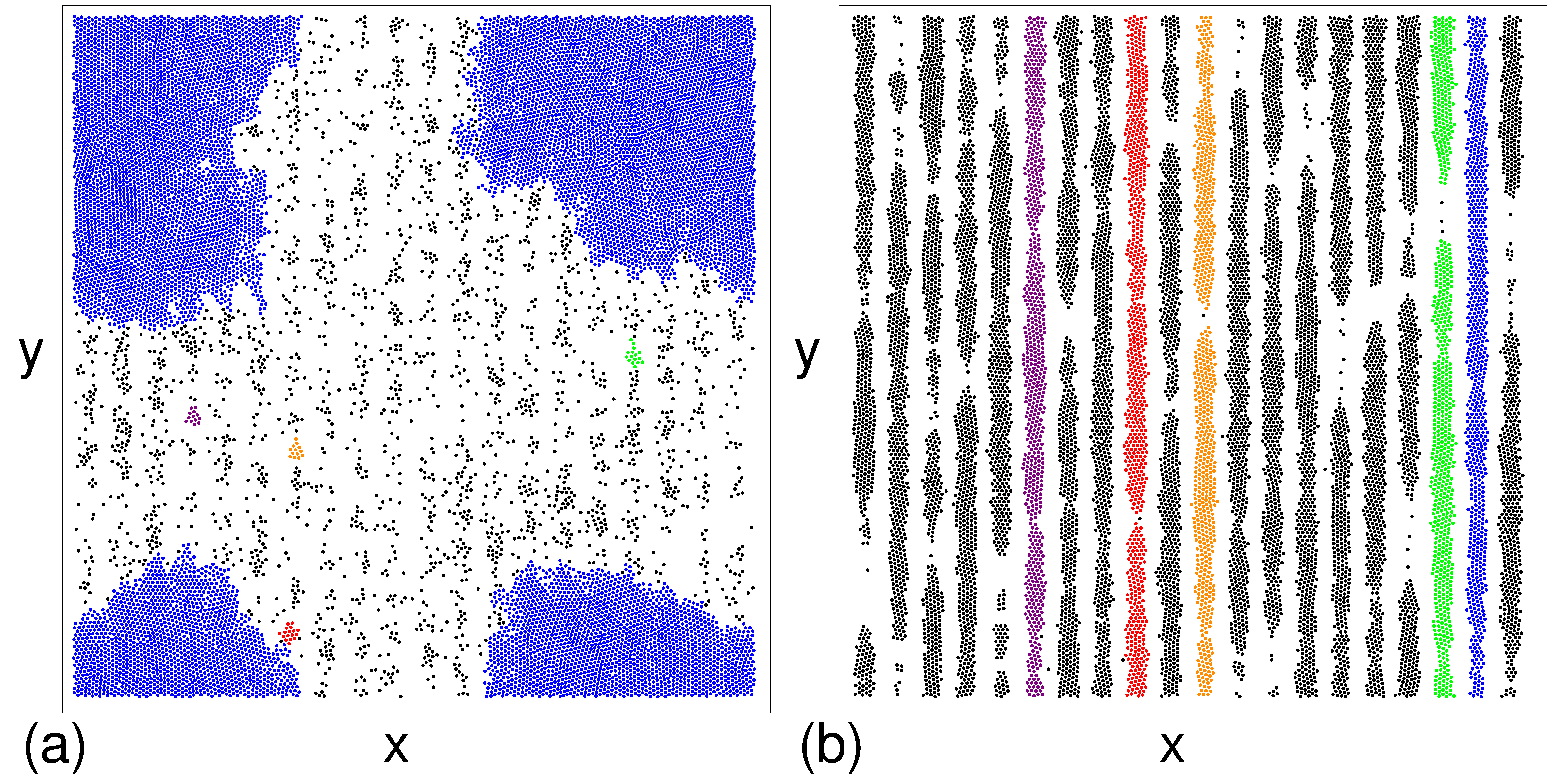}
\caption{ The real space positions of the active disks for the system in Fig.~\ref{fig:3}(a,b)
  with $v_{w} = 0.6$.
  (a) At $A_{s} = 0.75$, a phase separated state appears.
(b) At $A_{s} = 2.5$, the disks are strongly localized in the substrate minima and move
with the substrate.
}
\label{fig:4}
\end{figure}

To highlight the correlation between the changes in
the transport and the onset of clustering, in
Fig.~\ref{fig:3}(a,b)
we plot $\langle V\rangle/v_{w}$ and the normalized
$C_{L}=\tilde{C}_L/N$ versus $A_{s}$ at a fixed value of  $v_{w} = 0.6$
from the system in Fig.~\ref{fig:2}.
Here
the cluster to non-cluster transition occurs at $A_{s} = 1.25$,
as indicated by the drop in $C_{L}$
which also coincides with a jump in $\langle V\rangle/v_{w}$.
For this value of $v_{w}$, a complete locking between the disks and the traveling
wave occurs for $A_{s} \geq 3.0$, where
$\langle V\rangle/v_{w} = 1.0$.
In Fig.~\ref{fig:3}(c,d) we plot $\langle V\rangle/v_w$ and $C_L$ versus $A_s$ for the
same system at
$v_{w} = 2.0$, where the cluster to non-cluster transition
occurs at a higher value of $A_{s} = 2.0$.
This transition again coincides with a sharp
increase in $\langle V\rangle/v_{w}$.
In Fig.~\ref{fig:4}(a) we show images of the disk configurations for
the system in Fig.~\ref{fig:3}(a,b) with $v_w=0.6$ at
$A_{s} = 0.75$,
where the disks form a cluster state,
while in Fig.~\ref{fig:4}(b), at $A_s=2.5$ in the same system,
the clustering is lost and the disks are
strongly trapped in the substrate minima, forming
chain like states that move with the substrate.
These results indicate that the clusters act as composite objects that
only weakly couple to the substrate.

\begin{figure}
\includegraphics[width=3.5in]{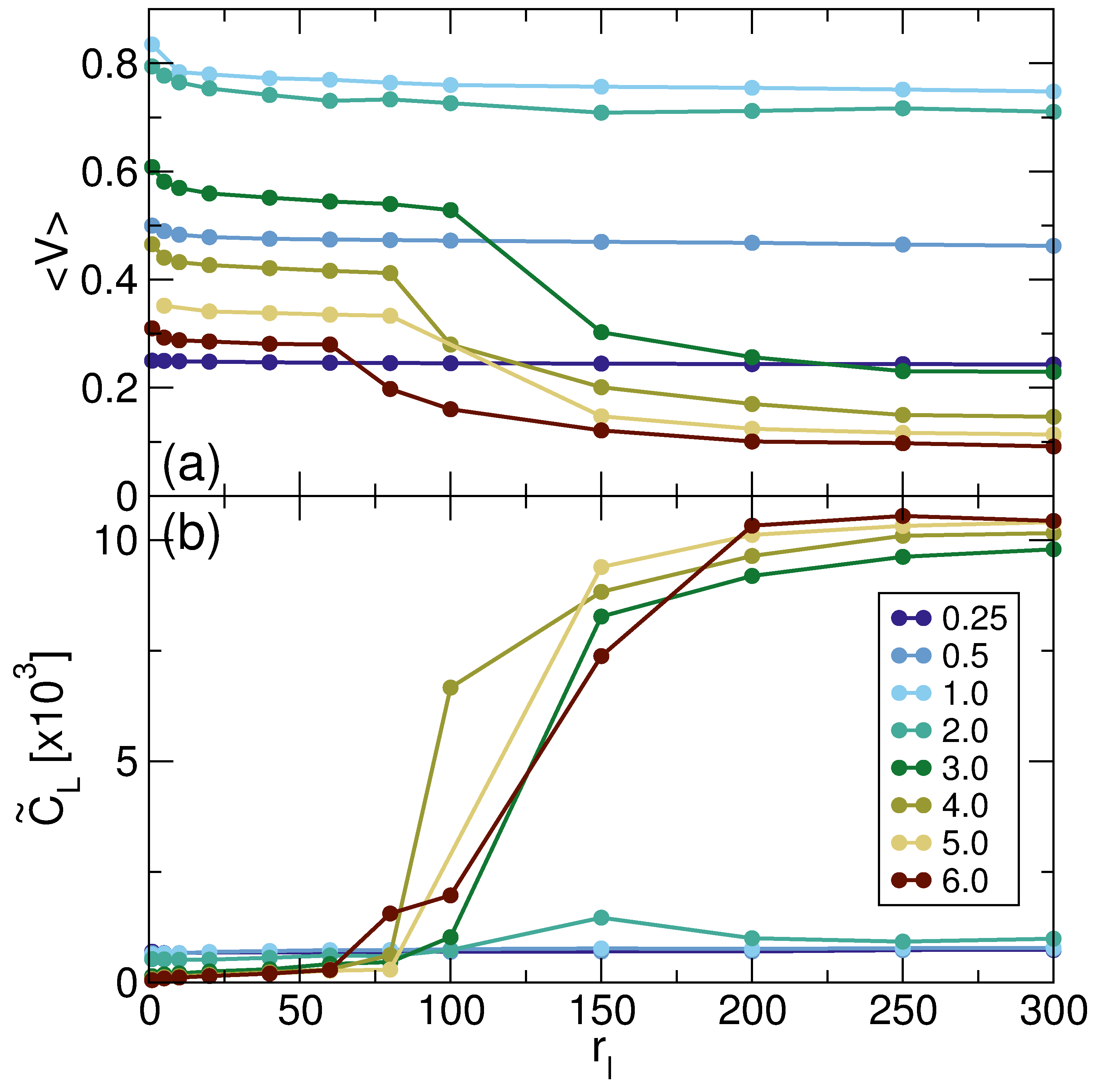}
\caption{(a) $\langle V\rangle$ vs $r_{l}$
  in samples with $A_{s} = 2.0$ and $\phi = 0.453$ for varied $v_{w}$ from
  $v_w=0.25$ to $v_w=6.0$.
(b) The corresponding $\tilde{C}_{L}$ vs $r_{l}$.}
\label{fig:5}
\end{figure}

\begin{figure}
\includegraphics[width=3.5in]{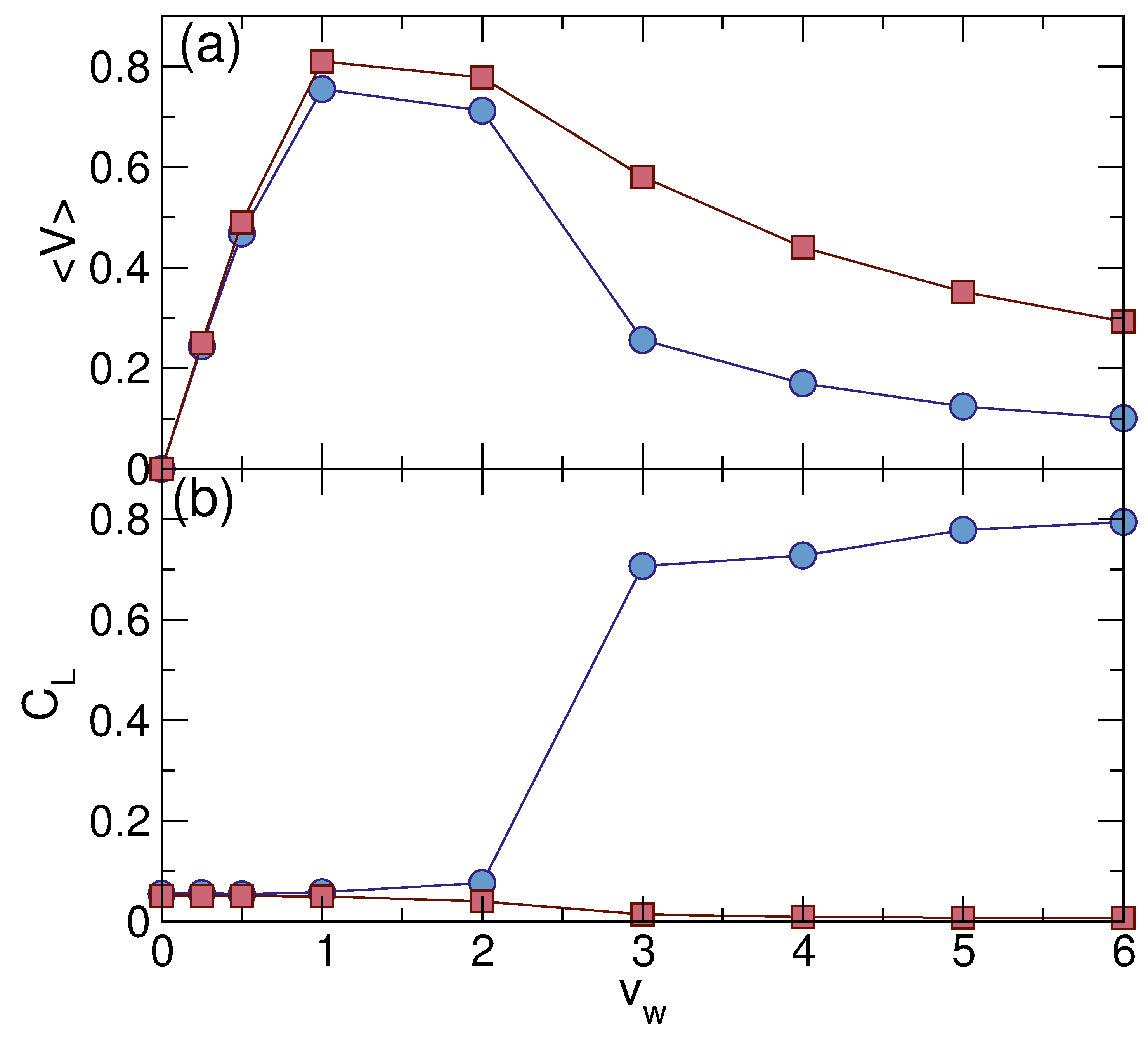}
\caption{A sample with $A_s=2.0$ and $\phi=0.453$
  for $r_{l} = 5$ (red squares) and $r_l=200$ (blue circles).
  (a) $\langle V\rangle$ vs $v_{w}$.
(b) $C_{L}$ vs $v_{w}$.
}
\label{fig:6}
\end{figure}

We next examine the case with a fixed
substrate strength of $A_{s} = 2.0$ and varied $r_{l}$.
Figure~\ref{fig:5}(a) shows
$\langle V\rangle$ versus $r_{l}$
for $v_{w}$ values ranging from $v_w=0.25$
to $v_w=6.0$, and Fig.~\ref{fig:5}(b) shows the corresponding $\tilde{C}_{L}$
versus $r_{l}$.
For $v_{w} < 3.0$ the system remains in a non-cluster state for all values of $r_{l}$, while
for $v_{w} \geq 3.0$ there is a transition
from a non-cluster to a cluster state
with increasing $r_l$ as indicated by the simultaneous drop in
$\langle V\rangle$ and increase in $\tilde{C}_{L}$.
In Fig.~\ref{fig:6}(a,b) we plot $\langle V\rangle$ and
$C_{L}$ versus $v_{w}$ at $A_{s} = 2.0$
for $r_{l} = 200$ and $r_l=5.0$.
The system is in a non-cluster state for all $v_w$ when
$r_{l} = 5.0$,
and there is a peak in $\langle V\rangle$ near $v_{w} = 1.0$,
while for $r_{l} = 200$ there is a transition to a cluster state
close to $v_{l} = 3.0$
which coincides with a drop in $\langle V\rangle$ that is much sharper than the decrease
in $\langle V\rangle$ with increasing $v_w$ for the $r_{l} = 5$ system.
In general, when $r_{l}$ is small, the net transport of disks through the sample is
greater than in samples with larger $r_l$.
The fact that the net disk transport varies with varying $r_l$ suggests
that traveling wave substrates could be used as a method for separating
different types of active matter, such as clustering and non-clustering species.

\begin{figure}
\includegraphics[width=3.5in]{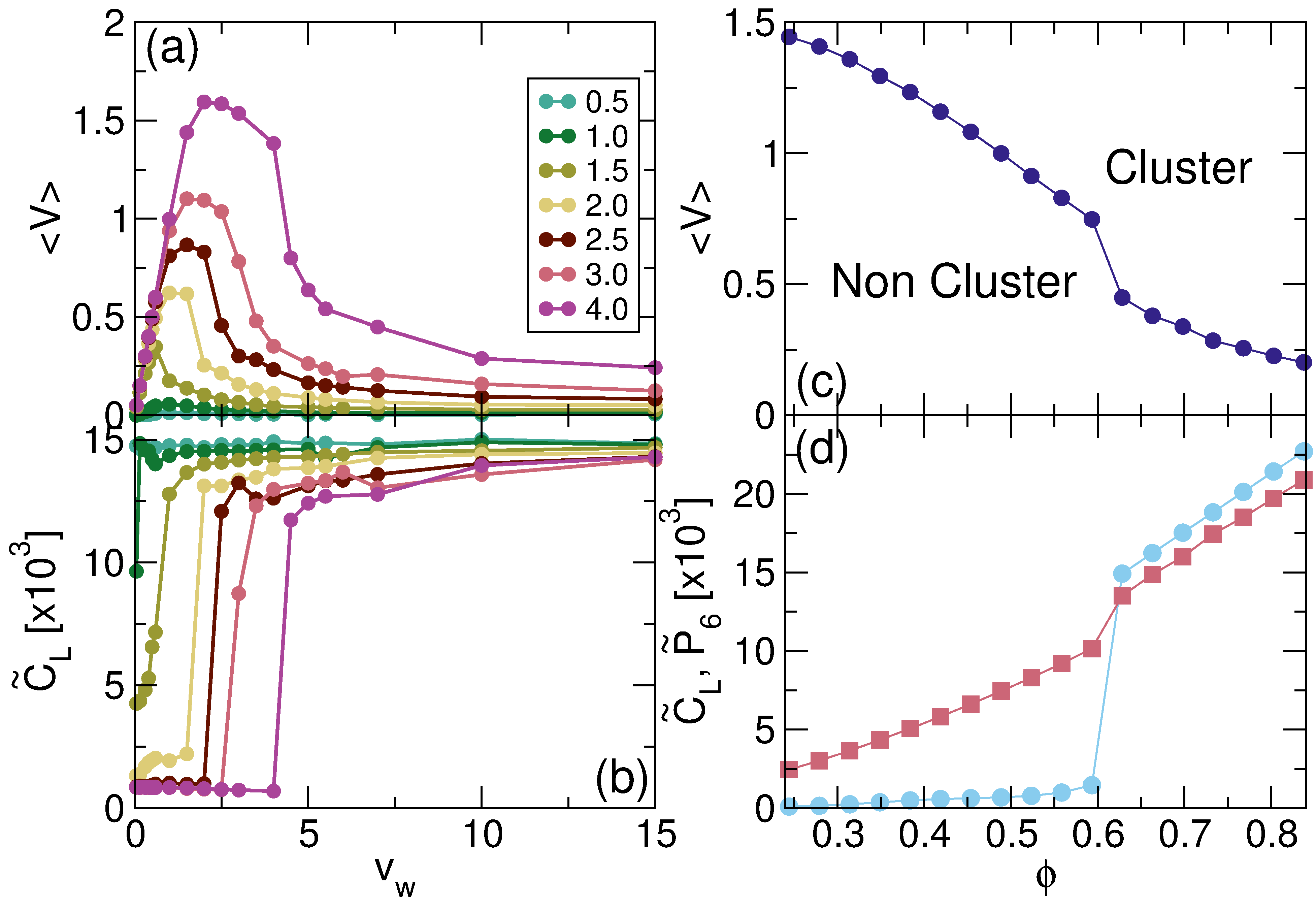}
\caption{
  (a) $\langle V\rangle$ vs $v_{w}$ at $\phi = 0.56$ and
  $r_{s} = 300$ for varied $A_{s} = 0.5$ to $A_s=4.0$.
(b) The corresponding $\tilde{C}_{L}$ vs $v_{w}$.
(c) $\langle V\rangle$ vs $\phi$ for $A_{s} = 2.5$, $r_{l} = 300$ and $v_{w} = 2.0$.
  (d) The corresponding
  $\tilde{C}_{l}$ (blue circles) and $\tilde{P}_{6}$ (red squares) vs $\phi$
  where the onset of clustering occurs
  near $\phi = 0.6$ at the same point for which there is a drop in
 $\langle V\rangle$ in panel (c).
}
\label{fig:7}
\end{figure}

When we vary the disk density $\phi$ while holding $r_l$ fixed, we find
results similar to those described above.
In Fig.~\ref{fig:7}(a) we plot
$\langle V\rangle$ versus $v_{w}$
at $\phi = 0.56$ for varied $A_{s}$ from $A_s=0.5$ to $A_s=4.0$, where we find
a similar trend in which
$\langle V\rangle$ increases with increasing wave speed
when the disks are strongly coupled to the substrate.
A transition to a cluster state occurs at higher $v_w$ as shown in
Fig.~\ref{fig:7}(b) where we plot $\tilde{C}_{L}$ versus $v_{w}$ for the same
samples.  The increase in $\tilde{C}_{L}$ at the cluster state onset
coincides
with a drop in $\langle V\rangle$.
In Fig.~\ref{fig:7}(c) we plot $\langle V\rangle$ versus $\phi$
for a system with fixed $v_{w} = 2.0$, $A_{s} = 2.5$, and $r_{l} = 300$, while
in Fig.~\ref{fig:7}(d) we show the corresponding
$\tilde{C}_{L}$ and $\tilde{P}_{6}$ versus $\phi$.
A transition from the
non-cluster to the cluster state occurs
near $\phi=0.6$, which correlates with a sharp drop in $\langle V\rangle$ and
a corresponding increase
in $\tilde{C}_{L}$ and $\tilde{P}_{6}$.

\begin{figure}
\includegraphics[width=3.5in]{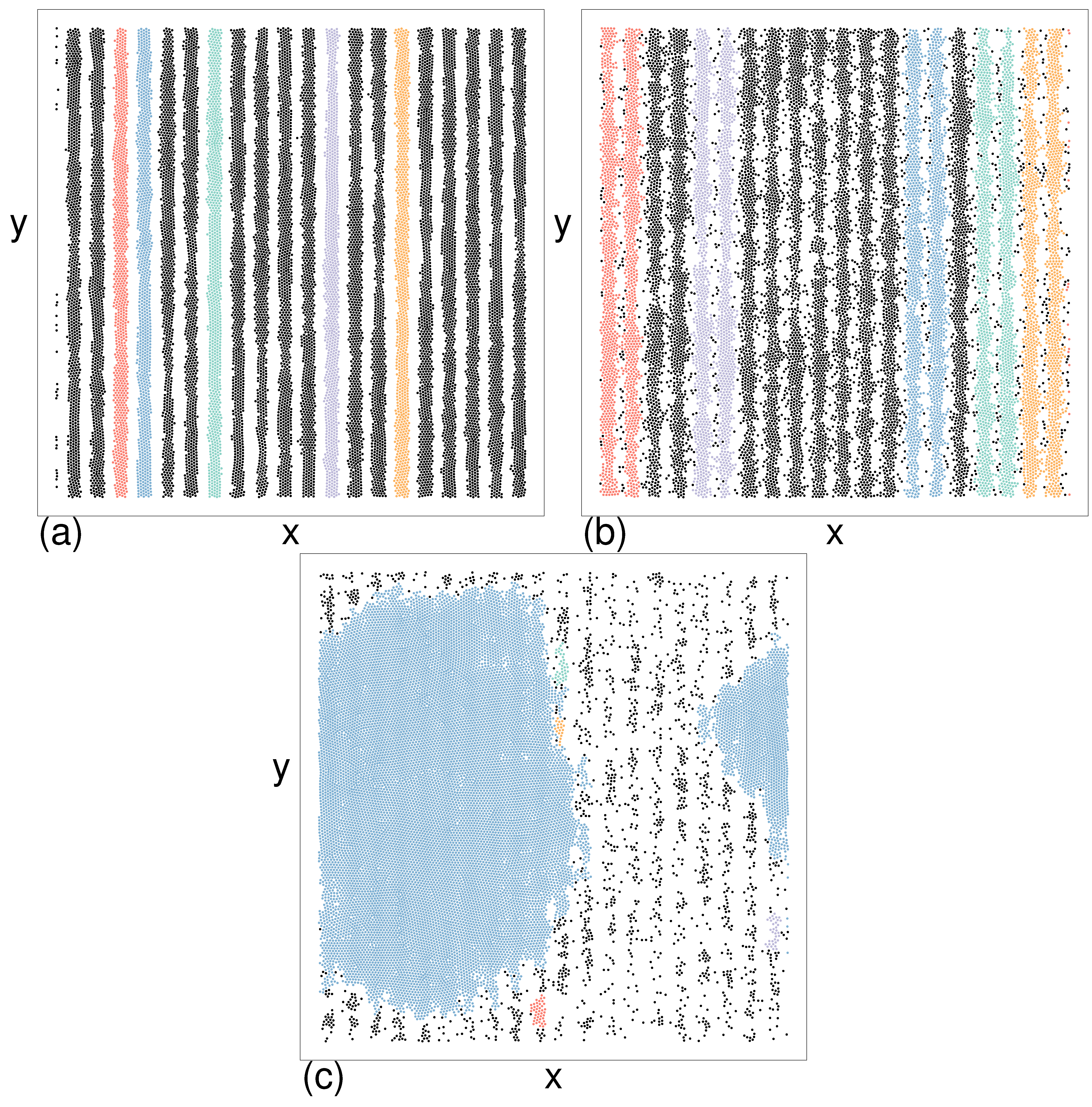}
\caption{
  The disk positions on the traveling wave substrate for the system in
  Fig.~\ref{fig:7}(a) at $\phi = 0.56$.  Colors indicate disks belonging to
  the five largest clusters.
  (a) Complete locking
  at $A_{s} = 4.0$ and $v_{w} = 1.0$,
  where the transport efficiency is $\langle V\rangle/v_{w} = 0.998$.
  (b) Partial locking
  at $A_{s} = 2.0$ and $v_{w} = 1.5$, 
  where $\langle V\rangle/v_{w} =  0.41$.
  (c) Weak locking
  at $A_{s} = 1.0$ and $v_{w} = 0.6$  
  with $\langle V\rangle/v_{w} = 0.078$.
}
\label{fig:8}
\end{figure}

In Fig.~\ref{fig:8}(a) we show the disk configurations from the system in
Fig.~\ref{fig:7}(a) at $A_{s} = 4.0$ and $v_{w} = 1.0$.
Here $\langle V\rangle/v_{w} = 0.998$,
indicating that the disks are almost completely locked with
the traveling wave motion
and there is little to no slipping of the disks out of the substrate minima.  
In Fig.~\ref{fig:8}(b), the same system at
$A_{s} = 2.0$ and $v_{w} = 1.5$
has a transport efficiency of $\langle V\rangle/v_{w} = 0.41$.
No clustering occurs but there are numerous disks that 
slip as the traveling wave moves.
At $A_{s} = 1.0$ and $v_{w} = 0.6$ in
Fig.~\ref{fig:8}(c)
there is a low transport efficiency of
$\langle V\rangle/v_w=0.078$.
The system forms a cluster state and smaller numbers of individual disks
outside of the cluster are transported by the traveling wave.

\section{Forward and Backward Cluster Motion}

\begin{figure}
\includegraphics[width=3.5in]{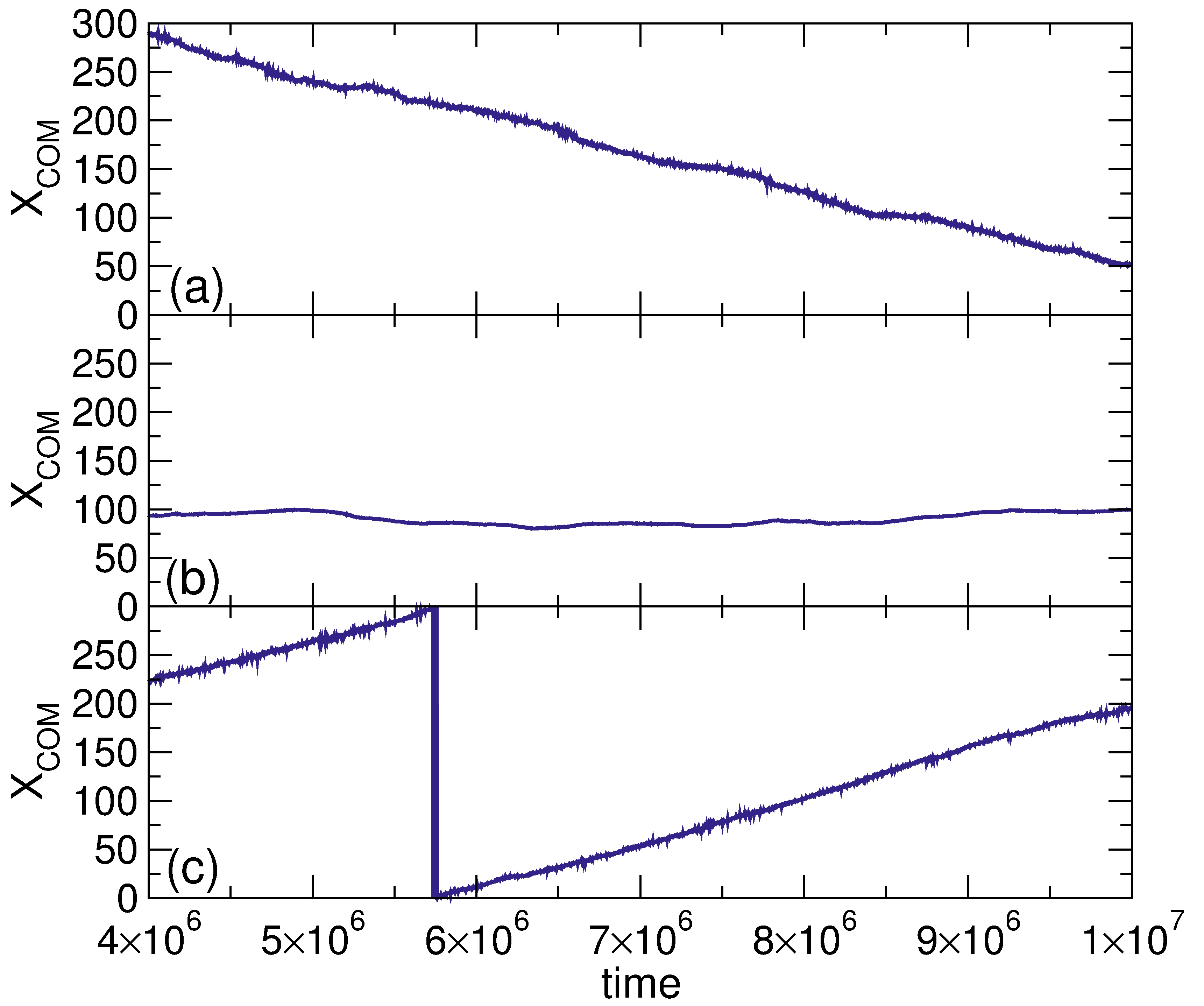}
\caption{
  The center of mass $X_{\rm COM}$ location of a cluster vs
  time in simulation time steps for a system with $\phi = 0.454$ and $r_{l} = 300$.
(a) At $A_{s} = 1.25$ and $v_{w} = 0.6$, the cluster moves in the negative $x$-direction,
  against the direction of the traveling wave.
  (b) At $A_{s} = 0.5$ and $v_{w} = 4.0$, the cluster
  is stationary.
  (c) At $A_{s} = 3.0$ and $v_{w} = 7.0$, the cluster moves in the positive
  $x$-direction, with the traveling wave.
  The dip indicates the point at which the center of mass
  passes through the periodic boundary conditions.
}
\label{fig:9}
\end{figure}

In general, we find that when the traveling wave is moving in the positive $x$-direction,
$\langle V\rangle > 0$; however, within the
cluster phase,
the center of mass motion of a cluster can be
in the positive or negative $x$ direction or the cluster can be almost stationary.
By using the cluster algorithm we can track the $x$-direction motion of the
cluster center of mass $X_{\rm COM}$ over fixed time periods,
as shown in Fig.~\ref{fig:9}(a) for a system with
$\phi = 0.454$, $r_{l} = 300$, $A_{s} = 1.25$, and $v_{w} = 0.6$.
During the course of $6\times 10^6$ simulation time steps
the cluster moves in the negative $x$-direction a distance of
$235$ units, corresponding to a space containing 16 potential minima.
Even though the net disk flow is in the positive
$x$ direction, the cluster itself drifts in the negative $x$ direction.
In Fig.~\ref{fig:9}(b) at $A_{s} = 0.5$ and $v_{w} = 4.0$,
the disks are weakly coupled to the substrate
and the cluster is almost completely stationary.
Figure~\ref{fig:9}(c) shows that at $A_{s} = 3.0$ and $v_{w} = 7.0$,
the cluster center of mass motion is now
in the positive $x$ direction, and the cluster
translates a distance equal to almost $20$ substrate minima
during the time period shown.
The apparent dip in the center of mass motion
is due to the periodic boundary conditions.

\begin{figure}
\includegraphics[width=3.5in]{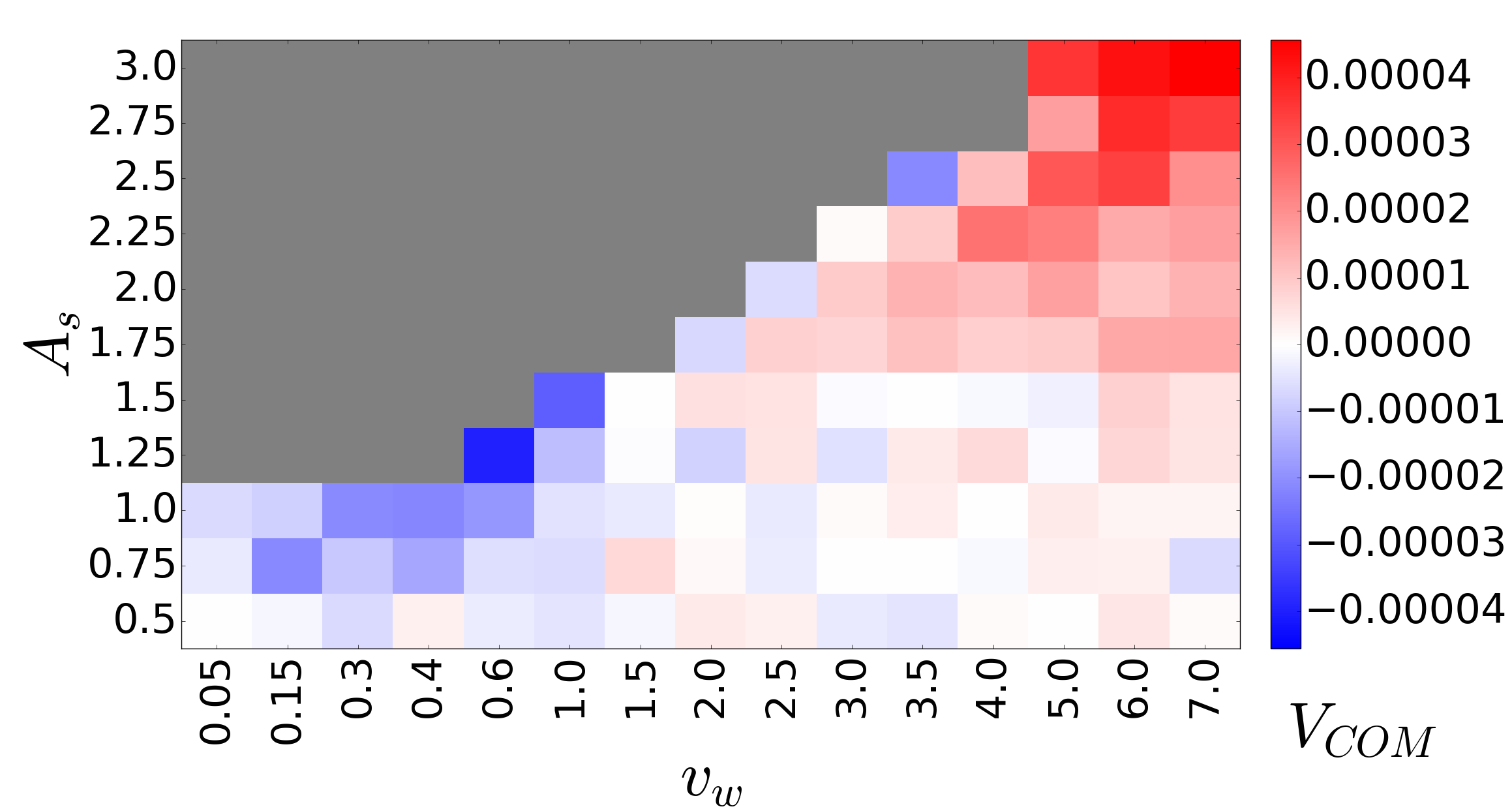}
\caption{
  Height field of the direction and magnitude of the center of mass motion $V_{\rm COM}$
  as a function of $A_s$ vs $v_w$ for the cluster
  obtained after $4\times 10^6$ simulation steps.
  The gray area indicates a regime in which there is no cluster formation.
}
\label{fig:10}
\end{figure}

\begin{figure}
\includegraphics[width=3.5in]{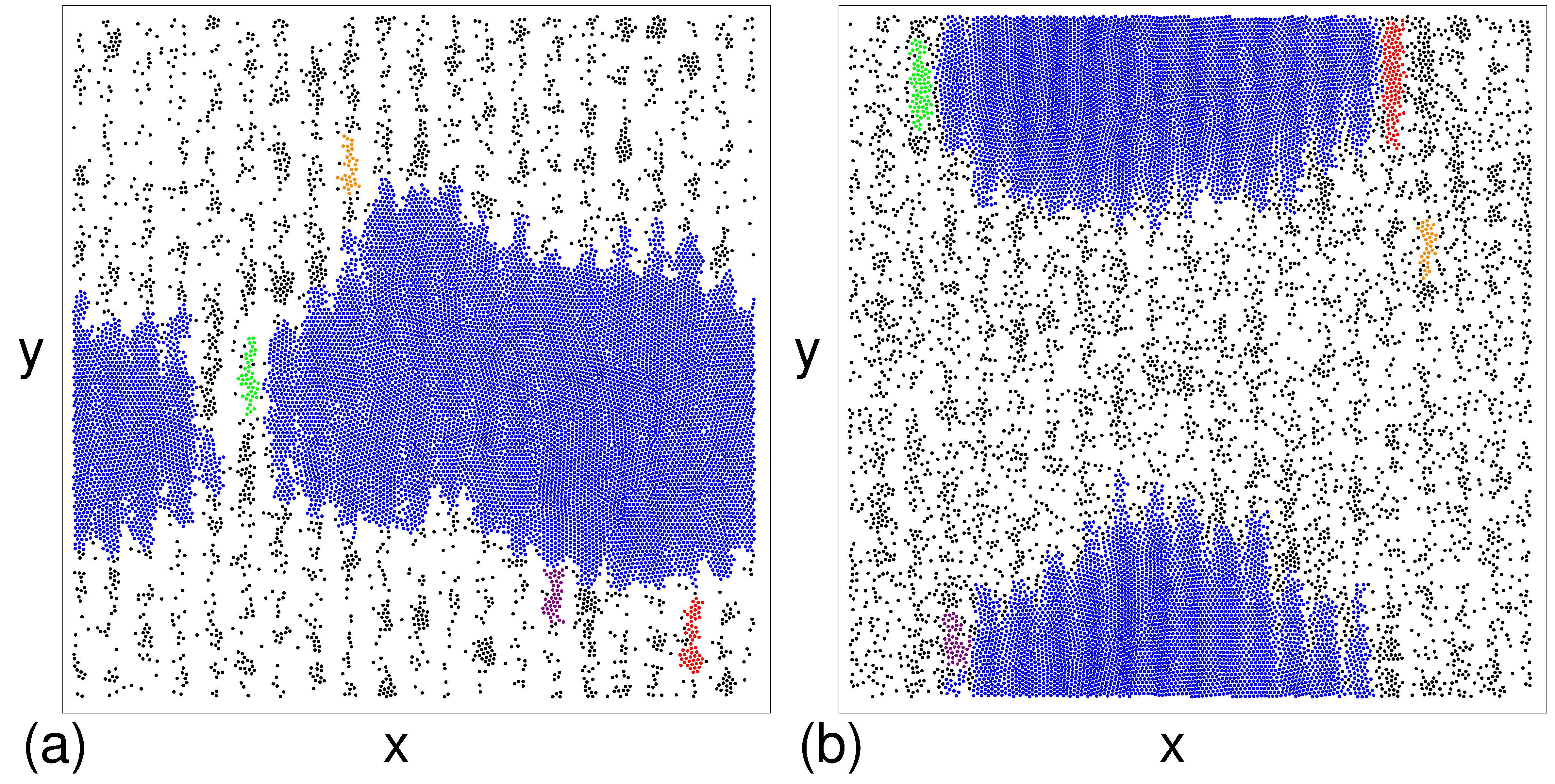}
\caption{ The disk positions for the system in Figs.~\ref{fig:9} and
  ~\ref{fig:10}.
  (a)
  At  $A_{s} = 1.0$ and $v_{w} = 0.4$,  the cluster drifts in the negative $x$-direction.
  (b)
  At $A_{s} = 3.0$ and $v_{w} = 5.0$, 
  the cluster drifts in the positive $x$-direction.
}
\label{fig:11}
\end{figure}

We have conduced a series of simulations and measured
the direction and amplitude $V_{\rm COM}$ of the center
of mass motion, as plotted
in Fig.~\ref{fig:10} as a function of $A_{s}$  versus wave speed
for the system in Fig.~\ref{fig:7}.
The gray area indicates a region in which clusters do not occur,
and in general we find that the negative cluster motion occurs
at lower wave speeds while the positive motion occurs
for stronger substrates and higher wave speeds.
There are two mechanisms that control the cluster center of mass motion.
The first is the motion of the substrate itself,
which drags the cluster in the positive $x$ direction, and
the second is the manner in which the cluster grows or shrinks
on its positive $x$ and negative $x$ sides.
At lower substrate strengths and low wave speeds,
the disks in the cluster are weakly coupled to the substrate
so the cluster does not move with the substrate.
In this case the disks can leave or
join the cluster anywhere around its edge; 
however, disks tend to join the cluster at a higher rate on its
negative $x$ side since individual disks, driven by the moving
substrate, collide with the negative $x$ side of the cluster and can become
trapped in this higher density area.  The positive $x$ side of the cluster
tends to shed disks at a higher rate since the disks can be carried
away by the moving substrate into the low density gas region.
The resulting asymmetric growth rate causes the cluster to drift in
the negative $x$ direction.
There is a net overall transport of disks in the positive $x$ direction due to the
large number of gas phase disks outside of the cluster region which follow
the motion of the substrate.
Figure~\ref{fig:11}(a) shows the
disk positions at $A_{s} = 1.0$ and $v_{w} = 0.4$, where the cluster
is drifting in the negative $x$ direction.     
For strong substrate  strengths, all the disks that are outside of the cluster become
strongly confined in the q1D substrate minima,
and the disk density inside the cluster itself starts to become modulated by the substrate.
Under these conditions, the cluster is dragged along with the traveling substrate
in the positive $x$ direction, as illustrated
in Fig.~\ref{fig:11}(b) for $A_{s} = 3.0$ and $v_{w} = 5.0$. 
These results
suggest that it may be possible for
certain active matter systems to 
collectively form a cluster state in order to 
move against an external bias
even when isolated individual particles on average move with the
bias.

\section{Summary}

We have examined run and tumble active matter disks interacting with
traveling wave periodic substrates.
We find that
in the non-phase separated state,
the disks couple to the traveling waves,
and that at the transition
to the cluster state,
there is a partial decoupling from the substrate and the net transport of disks by
the traveling wave  is strongly reduced.
We also find a transition from a cluster
state to a periodic quasi-1D liquid
state for increasing substrate strength,
as well as a transition back to a cluster state
for increasing traveling wave speed.
We show that there is a transition
from a non-cluster to a cluster state as a function of increasing
disk density which is correlated with a drop in the net disk transport.
Since disks with different run times drift with different velocities,
our results indicate that traveling wave substrates
could be an effective method for separating active matter particles with different mobilities.
Within the regime in which the system forms a cluster state,
we find that as a function of wave speed and substrate strength,
there are weak substrate regimes where the center of mass of the
cluster moves in the opposite direction from that of the traveling wave,
while for stronger substrates,
the cluster center of mass moves in the same direction as the traveling wave.
The reversed cluster motion occurs due to the
spatial asymmetry of the rate at which disks leave or join the cluster.
This suggests that collective clustering could be an effective method
for forming an emergent object that
can move  against gradients or drifts even
when individual disks on average move with the drift.

\acknowledgments
This work was carried out under the auspices of the 
NNSA of the 
U.S. DoE
at 
LANL
under Contract No.
DE-AC52-06NA25396.

\end{document}